\documentclass[american,english,final,1p,times,sort&compress]{elsarticle}
\usepackage[latin9]{inputenc}
\usepackage{float}
\usepackage{amssymb}
\usepackage{graphicx}
\usepackage{url}
\usepackage{lineno}

\makeatletter

\providecommand{\tabularnewline}{\\}

\usepackage[american,english]{babel}
\usepackage{color}
\usepackage{soul}

\def \etal {{\em et al.}}

\sethlcolor{yellow}

\makeatother

\usepackage{babel}

\journal{Nuclear Instruments and Methods A}

\begin{document}
\begin{frontmatter}

\title{KATANA - a charge-sensitive triggering system\\ for the S$\pi$RIT
experiment}

\author[IFJ,UJ]{P.~Lasko}
\author[UJ]{M.~Adamczyk}
\author[UJ]{J.~Brzychczyk}
\author{P.~Hirnyk}
\author[IFJ]{J.~{\L}ukasik}
\author[IFJ,email]{P.~Paw{\l}owski}
\author[UJ]{K.~Pelczar}
\author[UW]{A.~Snoch}
\author[UJ]{A.~Sochocka}
\author[UJ,dec]{Z.~Sosin}
\author[MSU,NSCL]{J.~Barney}
\author[NSCL]{G.~Cerizza}
\author[MSU,NSCL]{J.~Estee}
\author[RIKEN]{T.~Isobe}
\author[SEUL]{G.~Jhang}
\author[KYOTO]{M.~Kaneko}
\author[RIKEN]{M.~Kurata-Nishimura}
\author[MSU,NSCL]{W.G.~Lynch}
\author[KYOTO]{T.~Murakami}
\author[NSCL]{C.~Santamaria}
\author[MSU,NSCL]{M.B.~Tsang}
\author[TSINGHUA]{Y.~Zhang}

\fntext[email]{Corresponding author. E-mail: piotr.pawlowski@ifj.edu.pl}

\fntext[dec]{Deceased}

\address[IFJ]{Institute of Nuclear Physics, Polish Academy of Sciences, Kraków, Poland}
\address[UJ] {Faculty of Physics, Astronomy and Applied Computer Science, Jagiellonian University, Kraków, Poland}
\address[UW]{University of Wroclaw, Wroc{\l}aw, Poland}
\address[MSU]{Department of Physics and Astronomy, Michigan State University, East Lansing, USA}
\address[NSCL]{National Superconducting Cyclotron Laboratory, Michigan State University, East Lansing, USA}
\address[RIKEN]{RIKEN Nishina Center, Wako, Saitama, Japan}
\address[KYOTO]{Department of Physics, Kyoto University, Kita-shirakawa, Kyoto, Japan}
\address[SEUL]{Department of Physics, Korea University, Seoul, Korea}
\address[TSINGHUA]{Tsinghua University, Beijing, China}

\begin{abstract}

KATANA - the Krakow Array for Triggering with Amplitude discrimiNAtion - has
been built and used as a trigger and veto detector for the S$\pi$RIT TPC at
RIKEN. Its construction allows operating in magnetic field and providing fast
response for ionizing particles, giving the approximate forward multiplicity and
charge information. Depending on this information, trigger and veto signals are
generated. The article presents performance of the detector and details of its construction. A simple phenomenological parametrization of the number of emitted scintillation photons in plastic scintillator is proposed. The effect of the light output
deterioration in the plastic scintillator due to the in-beam irradiation is
discussed.

\end{abstract}

\end{frontmatter}

\section*{Introduction}

The symmetry term in the Nuclear Equation of State (NEoS) is being
recently intensively investigated \cite{tsang12,bali13,horowitz}. The experimental
efforts are focused on constraining the stiffness of the symmetry
energy, especially in the range of higher densities \cite{Russotto_PLB, Russotto_PRC}.
One of the proposed methods consists in getting the information on
momentum distribution of charged pions and isotopically-resolved light
charged particles (LCP), $Z\leq3$, emitted in the central heavy-ion (HI)
collisions \cite{BaoAn_PRL88,Rizzo_PRC72,Hong_Danielewicz}. Realization
of this kind of measurements inspired the construction of the Time
Projection Chamber (TPC) called S$\pi$RIT (the SAMURAI Pion-Reconstruction
and Ion-Tracker) \cite{Shane_NIMA784,genie16}. It is designed to
operate inside the vacuum chamber of the SAMURAI super-conducting spectrometer \cite{Kobayashi_NIMB_317}
at Radioactive Isotope Beam Factory (RIBF) of RIKEN \cite{Yano_NIMB261}. 

The experiments with the S$\pi$RIT TPC at RIKEN are up to now focused on investigation of the central and semi-central
collisions at energies of about 300 MeV per nucleon, where nuclear matter densities can reach up to $\sim$2$\rho_{0}$.
The target is located about 3 mm upstream of the entrance field cage window inside
the S$\pi$RIT TPC, so that the HI beam passes through the
chamber. To obtain sufficiently large gas amplification for pions
and light charged particles, a high electric field is required inside
the chamber. The ionization produced by the HI beam would produce an
excessive space charge leading to field distortion and could bring a risk
of a damage: the charge produced by gas ionization could exceed the safe limit
for the pad planes of the TPC. For this reason, a gating grid wire
plane was mounted in front of the pad plane \cite{Shane_NIMA784,Tangwancharoen}. 
The grid is normally 'closed' to keep the detector off until a desired
collision occurs. When it happens, the opening of the gating
grid can occur quickly (in $\sim$350~ns). 
Such operation should be performed whenever an incoming projectile
breaks up into much smaller fragments.

To produce a logic signal to open the gating grid when a desired event occurs, the KATANA detector, located downstream of the TPC exit window, was constructed. It is designed to play a double role: to produce a minimum bias or
majority trigger, and to provide a veto signal whenever a beam particle
or a fragment heavier than $Z\simeq20$ has passed through the chamber.
To fulfill the requirements, the wall has been constructed of two
parts, a Veto and a Trigger array. The KATANA-Veto part consisting
of 3 thin (1 mm thick) plastic-scintillator paddles with the middle
one centered on the beam, has been designed to produce a veto signal
for heavy fragments. The KATANA-Trigger array, consisting of 12 thicker
(10 mm thick) paddles, arranged on both sides of the beam, has been
designed to produce a forward multiplicity trigger. 

As the whole detector is expected to work inside the SAMURAI dipole magnet,
the apparatus has to be insensitive to magnetic fields up to about $0.5$-$1.0$
T. For this reason the light produced in the plastic scintillators is read-out
by silicon photomultipliers. All the materials used for the construction
are non-magnetic. 

Though the KATANA detector was designed as a triggering device for the S$\pi$RIT TPC, it may be used in another experimental context as well. Its versatile, modular and reconfigurable construction makes it easy to adopt other experimental tasks. Currently it is under preparation to work at Bronowice Cyclotron Center in Krak\'ow \cite{CCB}.

This paper will describe the technical details of the
KATANA wall construction and of the accompanying electronics. The performance
of the detector will also be discussed. In addition, some concepts concerning the modeling of light propagation in scintillators will be presented.

\section{Light propagation in thin plastic scintillators}

In order to make a realistic simulation of the light propagation in a plastic 
scintillator some basic ingredients are needed. One of the most important is the
mean number of the scintillation photons produced due to the excitation of
molecules along the particle track in the scintillator. The process depends in a
non-linear way on a specific energy loss of a given charged particle or fragment
and on its charge and mass \cite{birks}. So far, both the experimental data and
the model descriptions of the number of scintillation photons produced due to 
the light and heavy ions in plastic scintillators are rather scarce. This
results in a lack of a realistic description of this process in GEANT4 
\cite{geant}, which is a basic physics simulation environment. To fill this
gap, a simple parametrization of the mean number of the scintillation photons
produced in a plastic scintillator by charged fragments as a function of their
energy, charge and mass has been devised. The details of the parametrization are
presented in the Appendix.

The parametrization has been implemented into the GEANT4 to generate the mean
numbers of scintillation photons emitted in the VETO paddle during the passage
of charged particles. The distribution of charged particles has been generated
using the UrQMD transport code \cite{UrQMD}, with a clustering procedure, for
the \textsuperscript{132}Sn+\textsuperscript{124}Sn reaction at $E_{LAB}=300$
AMeV. The generated numbers of emitted photons as a function of particle type
and its kinetic energy have been used in a Monte Carlo code simulating the
light propagation in a scintillator. The real numbers of emitted photons are too
large to be followed in a Monte Carlo simulation. For this reason the numbers of
scintillation photons served here only to rescale a fixed number of simulated
test photons. This procedure allowed to limit the calculation time and to
obtain the realistic distributions at the end of the simulation. 

In the simulation we considered the propagation of the light generated in a
plastic scintillator of a size of 400$\times$100$\times$1 mm$^{3}$. For each
charged particle hitting the VETO paddle a point-size light source inside the
scintillator was assumed. The test photons were supposed to be emitted randomly
from this point with an isotropic distribution. Each simulated test photon has
been characterized by a weight, an emission time and a wavelength. The weight
was here a measure of probability of photon absorption. It was initialized to
one at the moment of emission, and reduced during the propagation, depending on
the attenuation length and the reflectivity of the walls. The test photon
wavelength and the emission time were attributed randomly, according to the
characteristics of the scintillator \cite{BC412}. 

The light was read out by a pixelated light sensor. Each simulated test photon
hitting the sensor area had a certain weight $W<1$. As the number of simulated
photons per particle was fixed (independently of the particle energy), the
weight was scaled by a factor corresponding to the real number of emitted
photons.  Then the actual number of photons hitting the sensor area was drawn
randomly from a Poisson distribution determined by the weight. The response of
the light sensor is governed by the photon detection efficiency (PDE) function
related to the wavelength of the photon \cite{MPPC}. If the photon is accepted
according to the PDE function, a randomly chosen pixel is activated. If the
pixel had not been activated before, it is marked as active, and produces an
electric signal of 5 ns fall time. The sum of all signals coming from all light
sources and all light sensors in a paddle was then digitized with a software
replica of a 14-bit, 500 MS/s Flash ADC.  The amplitude of the digitized pulse
was considered as a final output of the detector.

\begin{figure}[htb]
\begin{centering}
\includegraphics[width=6cm]{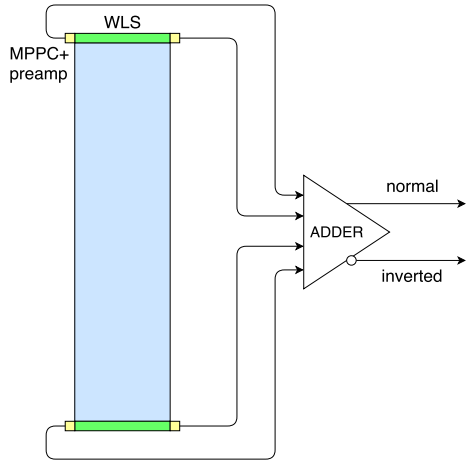} 
\par\end{centering}
\caption{Schematic view of a VETO paddle}
\label{fig:VETO_paddle}
\end{figure}

In the simulations a variety of light readout methods were considered. The best
found solution is schematically presented in Fig. \ref{fig:VETO_paddle} (see Sec. \ref{sec:technical} for more information on technical details). The light is first collected along two shorter edges of the plastic bar by a
wavelength shifting fiber (WLS) \cite{BCF92}. The WLS fiber is a special
plastic absorbing light and re-emitting it isotropically in another spectral
length. On both ends of the fiber the light sensors are mounted. This solution
was found to be less sensitive to the hitting-particle position than in the case
of light sensors coupled directly to the edge of the scintillator. 

\begin{figure}[h!]
\begin{centering} 
\includegraphics[width=8cm]{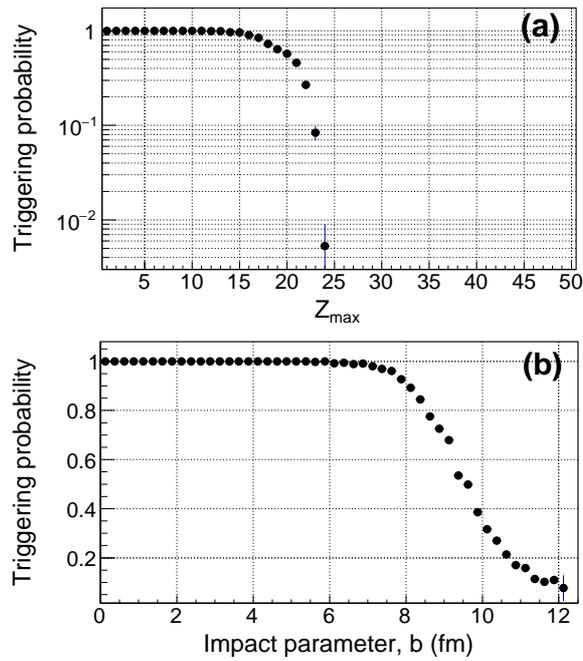}  
\par
\end{centering}

\caption{Simulated trigger efficiency of the KATANA detector: a) acceptance for
the charged fragments hitting the VETO paddles; b) distribution of the accepted
impact parameters.}

\label{fig:sim_efficiency}
\end{figure}

The main issue studied in the simulation was the efficiency
of charge filtering basing on the light signals generated by heavy
fragments in the VETO paddles. 
Figure \ref{fig:sim_efficiency} shows the predicted performance of the trigger/veto
system. The events generated by the UrQMD model were filtered by the
condition that at least one charged particle hits one of the 12 thick
(triggering) paddles and the VETO paddles do not produce the signal
higher than the value corresponding to $Z\sim20$. The figure presents the efficiency
of this triggering system as a function of the charge of the particle
hitting the VETO paddles (a), and as a function of the impact parameter
(b). 

The upper panel shows a very good performance of veto paddles in cutting-off the
heavier fragments using the amplitude threshold. Obviously, the most peripheral
collisions are cut-off by the VETO signal. This strong influence on impact
parameters accepted by the trigger is displayed in the lower plot.

\section{\label{sec:technical}Technical design of the detector}

\begin{figure}[htb]
\begin{center}
\includegraphics[width=10cm]{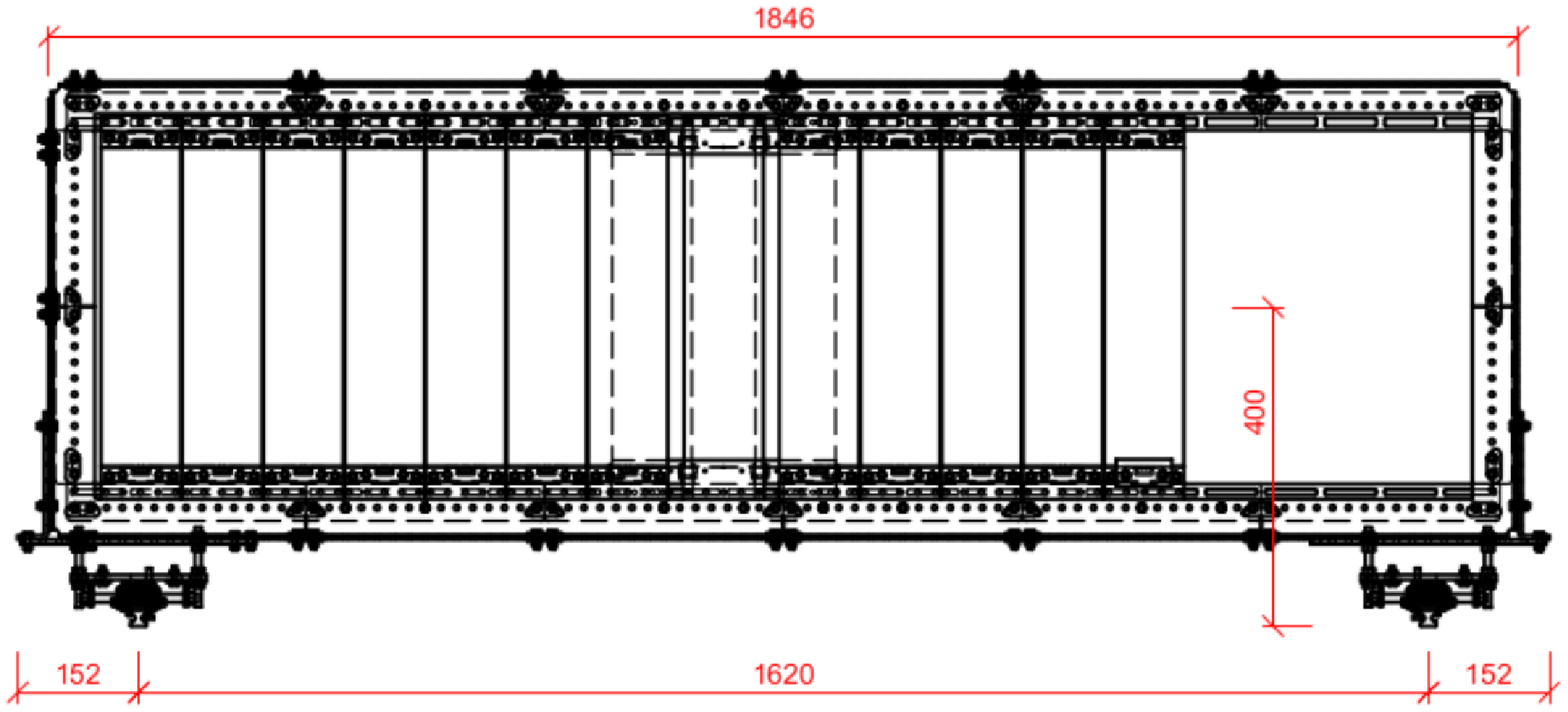} 
\includegraphics[width=10cm]{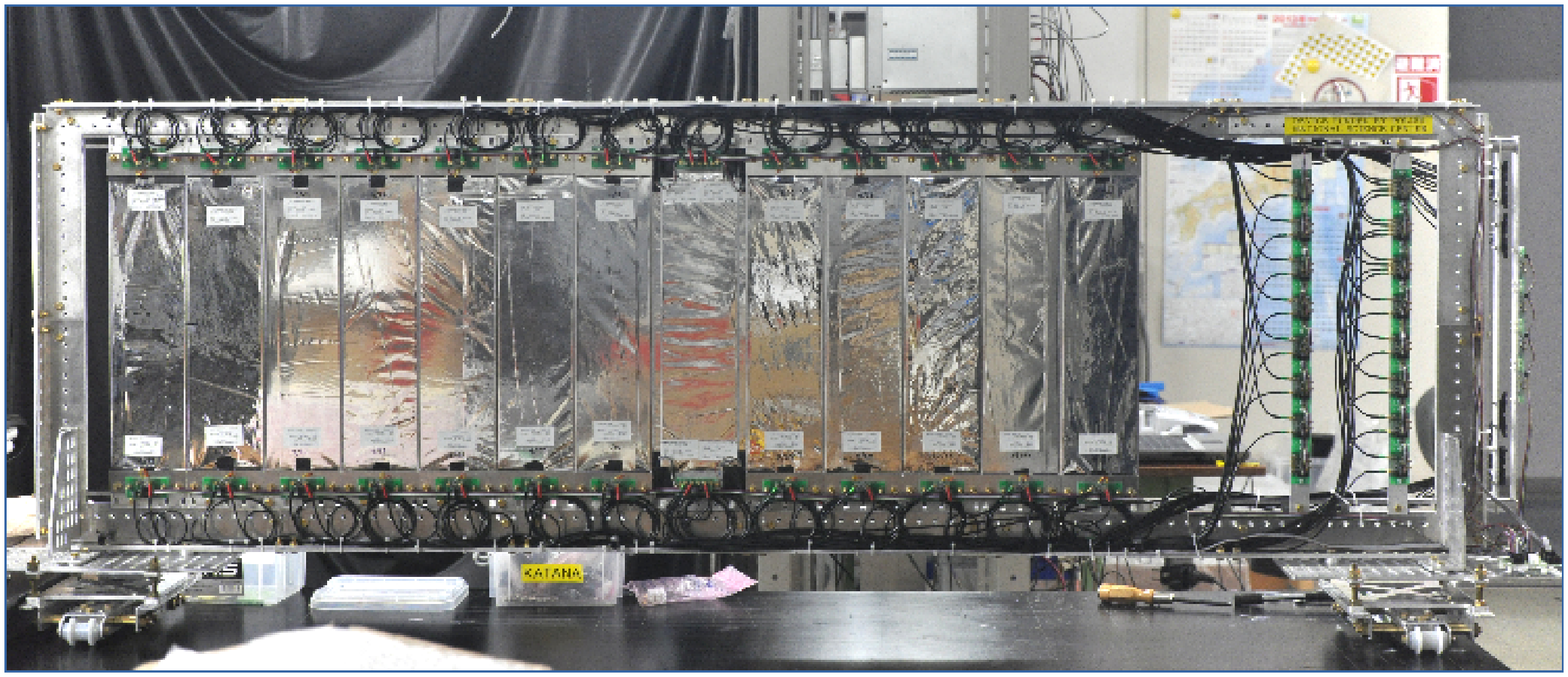}
\end{center}

\caption{Top panel: a technical drawing of the KATANA detector with the external
dimensions in mm. Bottom panel: a photo of the detector before mounting it inside the SAMURAI vacuum chamber.}
\label{fig:katana_overview}

\end{figure}

A technical design of the KATANA detector is presented in the upper part of Fig.
\ref{fig:katana_overview} while the bottom part presents its realization. Fifteen scintillators, each 10 cm wide and 40
cm high, are mounted on a rectangle, aluminum frame.  Each scintillator is
wrapped in a sheet of a light-reflective, metalized Kapton foil. Three veto
plastic bars, marked with dashed lines, (size $400\times100\times1$ mm$^{3}$,
made of BC-404) are placed in the middle of the frame. Twelve trigger plastics
($400\times100\times10$ mm$^{3}$, made of BC-408) are placed unevenly on both
sides of the beam (7 on the left and 5 on the right side) to account for bending
of the charged particles in the magnetic field. 

\begin{figure}[htb]
\begin{centering}
\includegraphics[height=5cm]{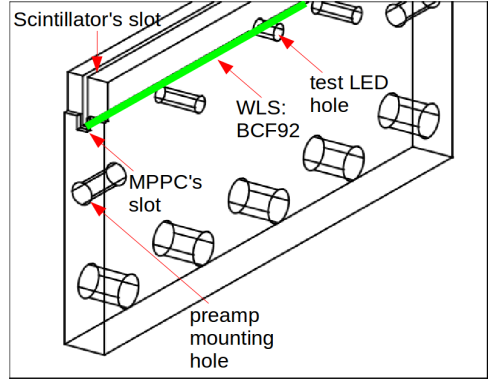}
\par\end{centering}
\caption{\label{fig:slot} Veto scintillator handle.}
\end{figure}

Each thick trigger paddle is fixed to the frame by two handles. The scintillation light is collected by two light sensors located in the middle of the
top and bottom edge of the plastic, and is read out directly without any
light guides. The thin veto paddles have more complex design
(see Fig. \ref{fig:slot}). In this case the light is read out by four light sensors. First, the light is collected along the plastic edge with the use of a WLS fiber (square cross section of $1\times1$ mm$^{2},$ type BCF-92). The WLS collects light from shorter sides of
the scintillator and propagates it to the light sensors mounted on both sides of
the fiber (see Fig. \ref{fig:VETO_paddle}). The fibers and the light sensors were mounted applying an optical grease as an interface. The main physical parameters of the plastics are presented in Tab. \ref{tab:plastic}.

\begin{table}[htb]
\centering{}\caption{\label{tab:plastic}Main physical properties of the plastic scintillators (BC-408 and BC-404) \cite{BC412} and WLS light guide (BCF-92)      \cite{BCF92}.}
\vspace{4mm}
\begin{tabular}{l|c|c|c}
 & BC-408 & BC-404 & BCF-92\tabularnewline
\hline 
\hline 
Emission maximum, $\lambda_{max}$ (nm) & 425 & 408 & 492\tabularnewline
Attenuation length, $L$ (cm) & 210 & 140 & $>$350\tabularnewline
Refractive index, $\eta$ & 1.58 & 1.58 & 1.42\tabularnewline
Light velocity, $v$ (cm/ns) & 18.9 & 18.9 & 21.1\tabularnewline
\end{tabular}
\end{table}

In both cases (thick and thin paddles), the Multi Pixel Photon Counter (MPPC)
devices from Hamamatsu \cite{MPPC} have been used as the light sensors. It is
worth mentioning that each MPPC is mounted directly on a preamplifier PCB to
minimize the noise level. The front-end electronics is completed with analog
adders providing the analog sum of signals coming from all sensors of a single
paddle. Thus, each paddle produces finally a single electric pulse. An inverted copy of the signal is available as well.

The frame of the KATANA detector is mounted on two trolleys, capable to
move on special rails used primarily to set-up the position of the
S$\pi$RIT TPC inside the SAMURAI vacuum chamber. The trolleys allow
also for left-right adjustment of the frame.

\section{Electronics}

\subsection{Light sensors}

As mentioned above, the light readout has been done using two types of MPPC. For the thick paddles the S12572-025P product was used.
Its photo-sensitive area of size $3\times3$~mm$^{2}$ consisted of
14400 parallelized pixel avalanche diodes. The thin paddles were read
out by the S12571-010P element containing 10000 pixels on $1\times1$~mm$^{2}$ area. Main parameters of the applied MPPCs are specified
in Tab. \ref{tab:MPPC_parameters}.

\begin{table}[htb]
\caption{\label{tab:MPPC_parameters}Main parameters of the applied MPPCs \cite{MPPC}.}
\vspace{4mm}

\begin{centering}
\begin{tabular}{l|c|c}
 & S12571-010P & S12572-025P\tabularnewline
\hline 
\hline 
Photosensitive area (mm\textsuperscript{2}) & 1$\times$1 & 3$\times$3\tabularnewline
Number of pixels & 10000 & 14400\tabularnewline
Typical operating voltage (V) & 69 & 68\tabularnewline
Gain & $1.35\times10^{5}$ & $5.15\times10^{5}$\tabularnewline
Peak sensitivity wavelength (nm) & 470 & 450\tabularnewline
Gain temp. coefficient (K\textsuperscript{-1}) & $1.6\times10^{3}$ & $8.2\times10^{3}$\tabularnewline
Peak photon detection efficiency (\%) & 10 & 35\tabularnewline
\end{tabular}
\par\end{centering}

\end{table}

\subsection{Preamplifier}

\begin{figure}[htb]
\begin{centering}
\includegraphics[width=11cm]{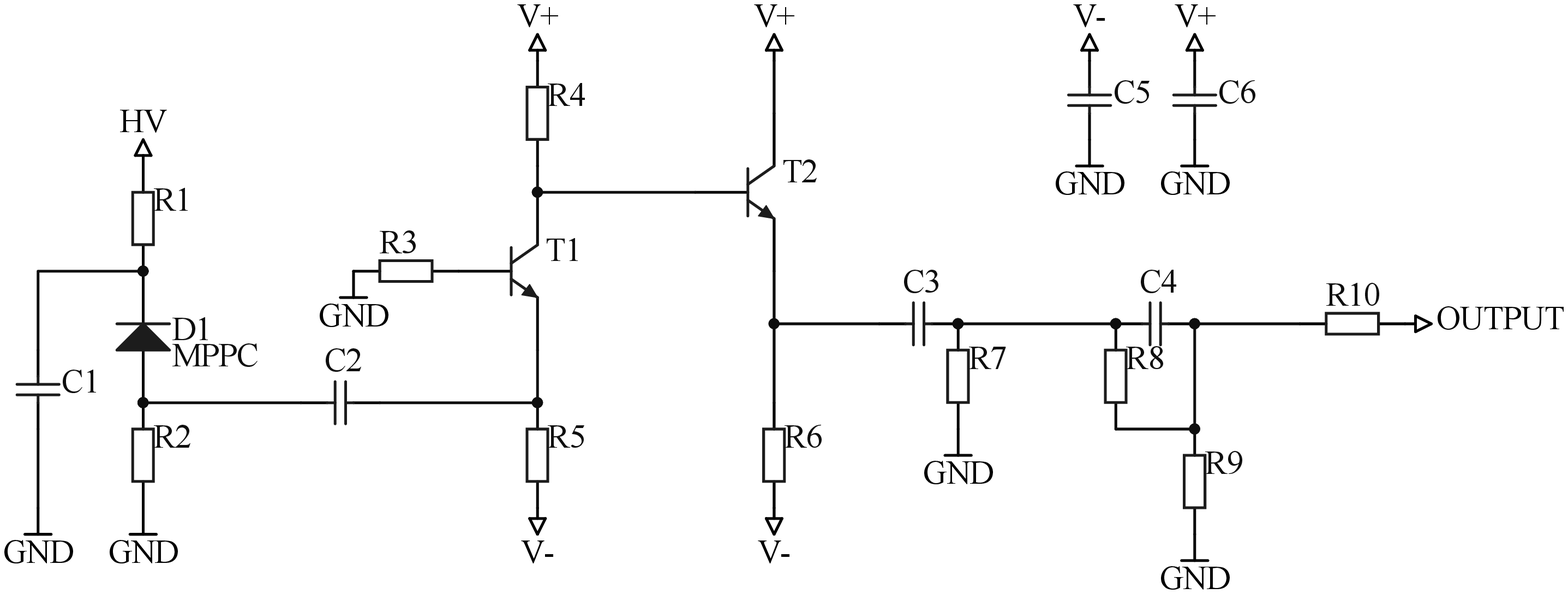}
\par\end{centering}

\caption{\label{fig:preamp}Schematic of the preamplifier for MPPC.}
\end{figure}

The circuit of  the preamplifier applied for MPPCs is shown in Fig.
\ref{fig:preamp}. It consists of a common base
and a common collector (emitter follower) amplifiers based on the
T1 and T2 transistors, respectively, and a simple pole zero cancellation
circuit based on the C3, R7 and R8 elements. The low voltage power
supplies are omitted in the schematic but it is obvious
that a high stability and low noise power supply is required.

\begin{table}[h!]

\caption{\label{tab:elements}The elements of the preamps applied for small and
big MPPCs. See Fig. \ref{fig:preamp} for the meaning of the symbols.}

\vspace{4mm}
\centering{}%
\begin{tabular}{c|c|c}
Element  & $1\times1$ mm$^{2}$ MPPC  & $3\times3$ mm$^{2}$ MPPC\tabularnewline
\hline 
\hline 
D1  & S12571-010P  & S12572-025P\tabularnewline
R1  & 1k$\Omega$  & 1k$\Omega$\tabularnewline
R2  & 60.4$\Omega$  & 1k$\Omega$\tabularnewline
R3  & 56$\Omega$  & 50$\Omega$\tabularnewline
R4  & 750$\Omega$  & 1.5k$\Omega$\tabularnewline
R5  & 750$\Omega$  & 5k$\Omega$\tabularnewline
R6  & 10 k$\Omega$  & 2 k$\Omega$\tabularnewline
R7  & 51$\Omega$  & 2k$\Omega$\tabularnewline
R8  & 4.7$\Omega$  & 0 \tabularnewline
R9  & -  & - \tabularnewline
R10 & 110$\Omega$  & 0 \tabularnewline
C1  & 100nF, 100V  & 100nF, 100V \tabularnewline
C2  & 10nF  & 10nF\tabularnewline
C3  & 100nF  & 100n F\tabularnewline
C4  & 100pF  & - \tabularnewline
C5, C6  & 1$\mu$F, 10V  & 1$\mu$F, 10V \tabularnewline
T1, T2  & BFR106  & BFR106 \tabularnewline
\end{tabular}
\end{table}

The MPPC (D1) is polarized by sufficiently high voltage (HV) provided
by HV power supply described below. High frequency signal pulses are
passing through C1 to the low impedance input of the common base amplifier.
This kind of amplifier was chosen because of its good characteristics
for high frequency operation.

A common collector amplifier was used to match the output impedance.
The R6 and R9 resistors also play a significant role in this respect.
The C2 capacitor cuts off the DC component. 

\begin{figure}
\begin{centering}
\includegraphics[width=8cm]{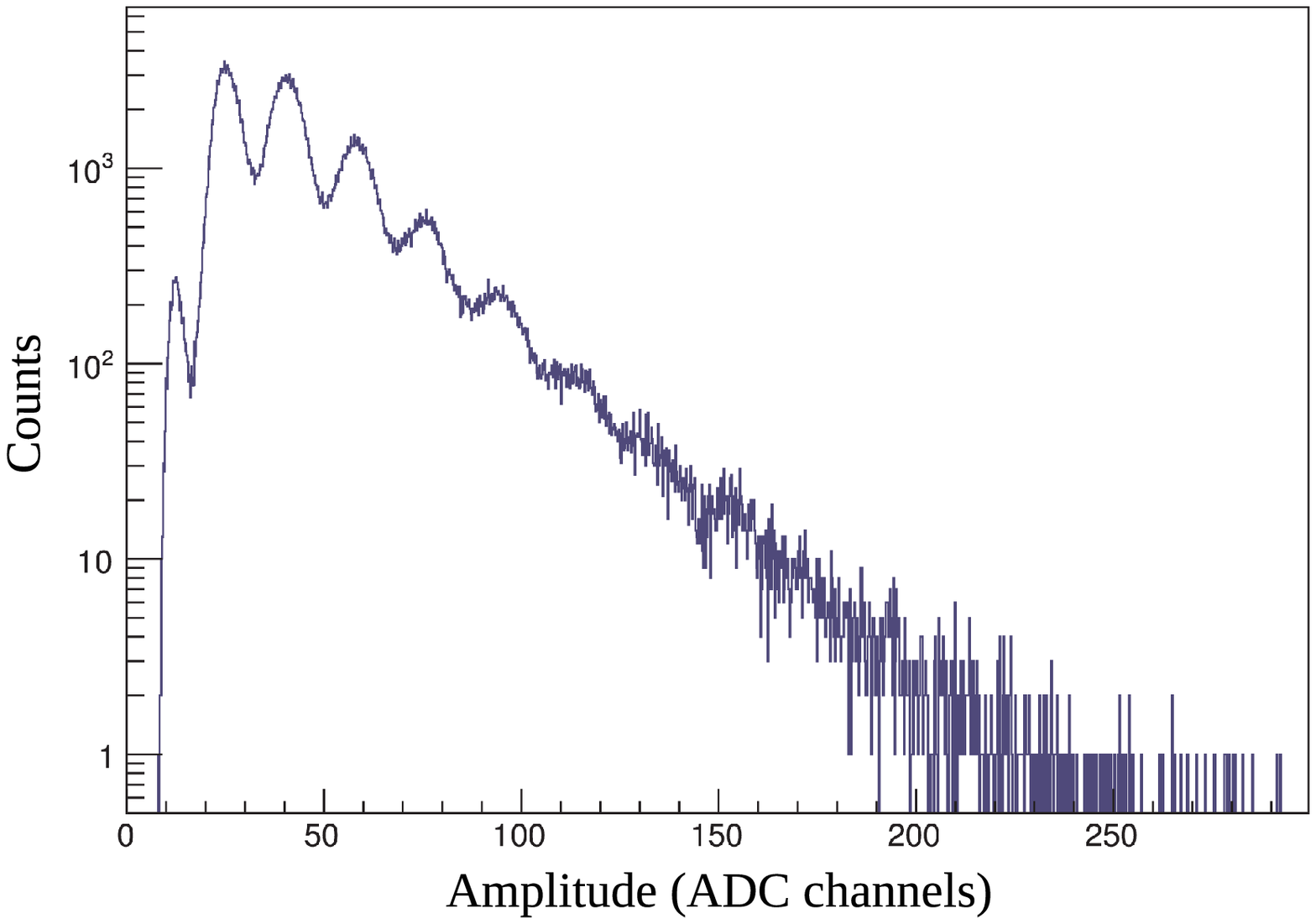} 
\par\end{centering}

\caption{\label{fig:photons}Single-photon amplitude resolution achieved for
thick paddles equipped with S12572-025P elements.}
\end{figure}

The preamplifier allowed to achieve a one-photon resolution in amplitude.
Figure \ref{fig:photons} shows the amplitude distribution obtained
from a thick paddle stored in a dark room. 
The peaks correspond to events with one, two and more photons. It clearly shows
sensitivity and amplitude resolution of the thick (trigger) paddles. Of course,
during the experiment, the thresholds were set-up much above this natural noise
to detect only light charged particles. On the other hand, the sensitivity of
thin (veto) paddles is much lower, as they should give response only for heavy
charged fragments.

The preamplifiers were optimized to obtain a gain sufficient to observe signals from
detectors in the range $0-2$ V. Because of different optical characteristics of
the small ($1\times1$~mm$^{2}$) and big ($3\times3$~mm$^{2}$) MPPCs used for
thick and thin paddles, the preamplifiers were prepared in two variants. Table
\ref{tab:elements} presents the parameters of the applied elements.

\subsection{High voltage power supply}

The MPPC elements need to be biased. This goal was achieved by designing a
dedicated HV power supply, satisfying the following special requirements: 10 mV
precision, low noise and low voltage ripple ($<1$~mV), $60-80$~V range of
regulation, 40 channels, remote control via Ethernet, multi-point temperature
control, current limiter and a feedback information (saturation/error). The device is
controlled by a STM32F103 micro-controller with ARM Cortex-M3 core. A separate Ethernet controller allows to create a system working in a server mode. The IP and port numbers for the communication are configurable. The interface for the DAC cards and HV cards was custom designed and allowed to connect those cards in parallel. Each card is provided with switches for setting the card address.

The HV supply is temperature sensitive. In order to monitor the temperature a DS18B20 element has been chosen as a built-in temperature sensor. This is a digital thermometer
with 12-bit ($0.0625^{\circ}$C) resolution and 1-Wire interface for
communication. Two such devices are installed inside the high voltage
power supply box. Additionally, six such sensors are mounted on the
frame of the KATANA detector to monitor the operating temperature
of the MPPCs. The temperature data are collected by the controller
and are available upon request from the server.

\subsection{Signal processing}

The amplified MPPC signals of each plastic scintillator are summed,
providing one analog signal for each scintillator. Each analog adder has normal
and inverted output. In total KATANA provides 15 normal (positive) signals and
15 inverted (negative) ones. The positive signals are sent to a Flash ADC board
(CAEN V1730) for control purpose. The negative signals are dispatched to the
leading edge discriminators and are compared to the threshold levels set by
remotely controlled DAC cards. The logic outputs of discriminators are then
analyzed by a logic circuit built on a FPGA board, providing majority trigger for the 12 triggering KATANA paddles (with the majority threshold adjustable between 1 and 12). The FPGA circuit takes also into account the other logic signals used in triggering. Several Gate\&Delay units are programmed in FPGA to synchronize the signals coming from different sources before performing logic operations (coincidence/anti-coincidence). KATANA multiplicity threshold, delay durations and gate widths can be remotely controlled via a RaspberryPi controller \cite{raspberrypi}.  

20-channel discriminator board, FPGA board, logic output buffers and RaspberryPi are integrated in a remotely controlled ``Trigger Box''. An integrated structure of the Trigger Box together with the fast and flexible FPGA logic allows to provide fast signals (within $<100$~ns) required by the TPC gating grid and by the acquisition system.

\section{Detector efficiency}

\begin{figure}[htb]
\begin{centering}
\includegraphics[width=12cm]{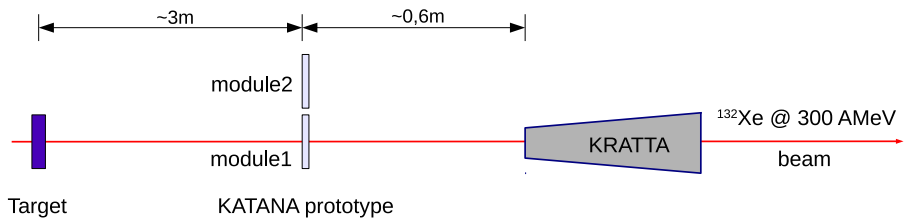} 
\par\end{centering}

\caption{Test run setup at HIMAC.}
\label{fig:himac_setup}
\end{figure}

The efficiency for HI detection and fragment discrimination is a crucial
characteristic of the KATANA-Veto paddles and has been tested using the 300 AMeV
$^{132}$Xe beams at HIMAC (the Heavy Ion Medical Accelerator in Chiba). The test
setup is depicted in Fig. \ref{fig:himac_setup}. 

\begin{figure}[htb]
\begin{centering}
    \includegraphics[width=8cm]{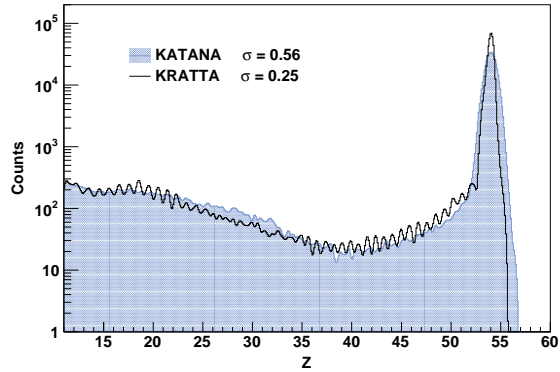}
\par\end{centering}

\caption{Charge resolution obtained at HIMAC with the KATANA prototype plastic
bar (hatched histogram) and with the KRATTA module (black line) for the
$^{132}$Xe fragmentation at 300~AMeV. Note the single charge resolution for the
KRATTA detector.}

\label{fig:KRATTA_charge}
\end{figure}

The prototype of the KATANA-Veto module was placed directly in the beam, in air.
A KRATTA detector module \cite{KRATTA} was mounted behind the veto paddle. The
KRATTA module is a triple telescope allowing to identify the charge of a hitting
particle. For the test, the gain of the KRATTA module was lowered to
increase its charge identification range. Using it as a reference takes advantage of  its high detection efficiency and its charge resolution. The charge
spectra collected by the KATANA-Veto prototype and the KRATTA module in
coincidence are shown in Fig. \ref{fig:KRATTA_charge}. In both cases the charge
resolution for the elastically scattered beam particles could be estimated. The
FWHM of the beam peaks amounts to about 0.6 and 1.3 charge units in the KRATTA
and the KATANA case, respectively. The former reflects the resolution of a 1 mm
thick silicon detector and the latter the resolution of 1 mm thick BC-404
plastic scintillator. In the case of the KRATTA detector the observed peaks correspond to resolution of single charges. 

\begin{figure}[htb]
\begin{centering}
\includegraphics[width=8cm]{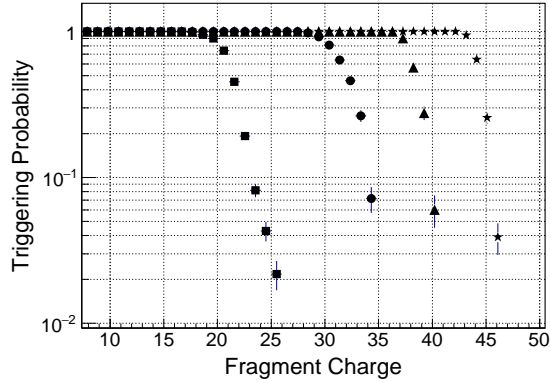} 
\par\end{centering}

\caption{Measured efficiency of a thin VETO paddle. Different symbols (squares, points, triangles and stars) correspond to increasing thresholds applied to the VETO signal amplitude.}
\label{fig:exp_efficiency}
\end{figure}

The efficiency of the veto module for the HI discrimination could be determined
by correlating the signal amplitude from the veto paddle with the fragment
charge number provided by the KRATTA module. Figure \ref{fig:exp_efficiency}
shows the detection efficiency of the VETO paddle as a function of the hitting
fragment charge for various threshold settings on the discriminator. The fragment charge selectivity of the KATANA veto paddle is clearly visible. We can also conclude that the simulated result presented in Fig. \ref{fig:sim_efficiency}a is realistic.

\section{Radiation damage}

\begin{figure}[htb]
\begin{centering}
\includegraphics[width=8cm]{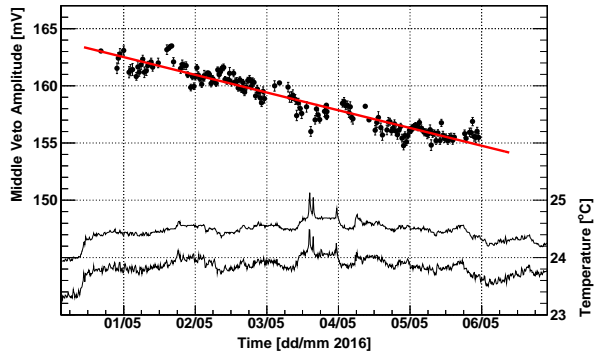}
\par\end{centering}

\caption{In-beam aging of a 1mm thick BC-404 plastic scintillator. The points
show the mean amplitudes obtained for the beam particles ($^{108}$Sn and $^{112}$ Sn
at 300 AMeV). The line shows a linear fit to the points. The histograms show the local
temperature variation at the top (upper histogram) and at the bottom (lower one)
of the middle Veto paddle.}

\label{fig:aging}

\end{figure}

As the veto paddles are exposed to the beam directly, a significant drop of
their efficiency of conversion of the absorbed energy into light may be expected
due to the radiation damage. During the S$\pi$RIT experiment a cocktail of beams
around $^{108}$Sn and $^{112}$Sn isotopes with an average intensity of $\sim$7.0 kHz
was applied. Even for such a low beam rate a decrease of light output is
expected after dozens of hours of exposure. The damage is reflected in the decrease
of the amplitude measured for the pure beam events. Figure \ref{fig:aging}
presents this effect as a function of the experimental time. The amplitude
decrease is about 1\% per beam day. On the other hand, no significant effect on
the width of  the beam peak was observed. 

It should be underlined that the data used for the radiation damage studies do
not come from a dedicated and optimized experiment. Rather, they are monitoring data taken with the use of the KATANA triggering system during the main S$\pi$RIT experiment. This may be the reason for a large scatter of the data points corresponding to the individual runs with possible changes in the beam conditions (position, composition, intensity). The two histograms at the bottom of Fig. \ref{fig:aging} show the variation of the local temperature measured at the top (upper histogram) and at the bottom (lower one) of the middle Veto bar. The temperature at the top was about 0.6 $^{\circ}$C higher because of the  proximity of the TPC electronics.
The sharp spikes around the middle of the beam time indicate the presence of the
experimenters in the vicinity of the sensors during the beam breaks. The
gain-temperature coefficient for the applied sensor (see Tab.
\ref{tab:MPPC_parameters}) implies the gain change of about 1\%/$^{\circ}$C.
Thus, the observed relative stability of the temperature near the MPPC sensors
excludes the temperature effect on the MPPC gain that might potentially
contribute to the observed amplitude variation.

\section*{Summary}

In summary, we have presented technical details and performance of the KATANA
array. The detector was designed to provide trigger/veto signal for the S$\pi$RIT
experiment at RIKEN. It was playing a significant role in the detection system,
protecting the TPC against a risk of damage by excessive ionization produced by HI
beams. Fast response, high veto and trigger efficiency, insensitivity to
magnetic fields, stability, portability and the possibility of remote control
are the main attributes of this array. 

\section*{Acknowledgments}

This work was supported by the Polish National Science Center (NCN), under contract Nos. UMO-2013/09/B/ST2/04064 and UMO-2013/10/M/ST2/00624 as a part of the S$\pi$RIT Project supported by the U.S. Department of Energy under Grant Nos. DE-SC0004835, DE-SC0014530, DE-NA0002923, US National Science Foundation Grant No. PHY-1565546 and the Japanese MEXT KAKENHI  (Grant-in-Aid for Scientific Research on Innovative Areas) grant No. 24105004. The authors are indebted to HIMAC and RIBF technical staffs for an excellent beam time.

\medskip{}

\section*{Appendix: Mean number of scintillation photons}

In order to obtain a realistic number of scintillation photons hitting the light
sensor in a simulation one needs first of all a realistic number of photons
produced along a particle track. The realistic number of photons hitting the
light sensor is particularly important when simulating the response of
pixelated light sensors. This is because the simulation has to provide an
optimal number of pixels assuring the required dynamic range of the device, or
in other words, the number of pixels sufficient for detection of the maximum
ionizing fragment before the sensor saturates. A fixed number of scintillation
photons per MeV, independent of the energy and of the ion identity, offered by
GEANT4 seems to be an oversimplification and was the main motivation to devise a
more realistic energy--light conversion formula.

The experimental data on the number of scintillation photons produced in plastic
scintillators are rather scarce. The most comprehensive  set of data on a
relative light output for various ions at intermediate energies is presented in
\cite{becchetti} and refers to the NE-102 scintillator. The authors provide
simple parameterizations of the light output as a function of the incident
energy and the mass and atomic numbers,  however these are very approximate.
Attempts to better describe the above data  are presented in
\cite{matsufuji99,menchaca09}. There, the models are able to  better describe
the data, however the differential form of the obtained parameterizations is
not best suited for Monte-Carlo simulations and energy calibrations. Instead,
we attempted to obtain a simple parametrization of the  light output assuming
the following formula:

\begin{equation}
N(E,A,Z) = a_{0} \left(
E^{a_{4}} - a_{1} A^{a_{6}} \ln(1+E^{a_{4}})-
a_{2} A^{a_{6}} Z^{a_{7}} \ln \left(1+\frac{E^{a_{4}}}
{a_{3} A^{a_{6}} Z^{a_{7}}}\right)\right)^{a_{5}}
\label{eqneaz}
\end{equation}

\noindent where $N$ is the mean number of scintillation photons produced in
a scintillator and $E$, $A$ and $Z$ are the energy in MeV, and the mass and
atomic numbers of an incident ion, respectively.

The following parameter values have been obtained by fitting the formula
(\ref{eqneaz}) to the relative light output data for the NE-102 scintillator:
$a_{0} = 89.435(\times 219)$, $a_{1} = 0.581$, $a_{2} = 0.0914$, $a_{3} = 0.29$,
$a_{4} = 0.34$, $a_{5} = 3$, $a_{6} = 0.193$ and $a_{7} = 2.762$. An additional
factor of 219 has to be applied to the overall scale factor, $a_{0}$, in order
to match the proton data for the NE-102 with the absolute numbers of
scintillation photons obtained for protons in the BC-400 scintillator
\cite{bc400}.

\begin{figure}
\begin{center}
    \includegraphics[width=8cm]{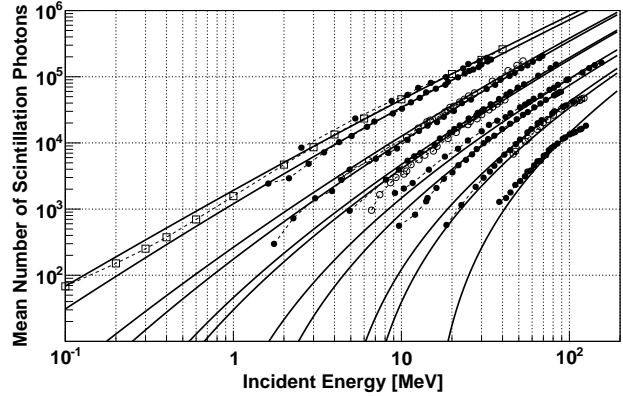}
\caption{The mean number of scintillation photons produced in a plastic
scintillator by charged particles of a given incident energy. The squares
represent the absolute numbers of photons emitted due to protons incident on a
BC-400 plastic scintillator \cite{bc400}. The open and filled circles represent
the experimental data of \cite{becchetti} for $^{1}$H, $^{2}$H, $^{3}$He,
$^{4}$He, $^{6}$Li, $^{7}$Li,  $^{12}$C, $^{16}$O, $^{32}$S, $^{40}$Ca and
$^{81}$Br from top to bottom, respectively. The original relative light output
for these data points has been multiplied by a factor of 219 to match the
absolute numbers of photons represented by the squares. The solid lines
represent the fits of the formula (\ref{eqneaz}) to the data. The dashed lines connect the data points for a given incident ion.}
    \label{fig:scint_phot}
\end{center}
\end{figure}

The form of Eq. (\ref{eqneaz}) has been inspired by the results of
\cite{parlog02} which, apart from the $\delta$-electron term, have been obtained
for the Birks' formula assuming the 0-th order Bethe-Bloch approximation:
$\Delta$E $\propto$ A Z$^2$/E. Its extension by introducing an additional
logarithmic term and applying exponents to the energy, number of photons, mass
and atomic numbers is purely phenomenological. The quality of the fit is
presented in Fig. \ref{fig:scint_phot}, together with the scaled experimental
data of \cite{becchetti} and with the data for protons from \cite{bc400}. The
$\chi^{2}$ of the fit amounts to about 0.8. The main motivation for devising the
above formula was to enable realistic simulations of the number of scintillation
photons produced in plastic scintillators, however its functional form (or its
inverse) may possibly also be used as an energy--light conversion formula for the energy calibration of plastic detectors. 

\end{document}